\documentstyle[multicol,eqsecnum,aps,prd,epsf]{revtex}
\begin{document}
 \def\rE{r_{\rm E}}
 \def\H0{H_{\rm 0}}
 \def\la{\lambda}
 \def\Msun{M_{\odot}}
 \makeatletter
 \newdimen\ex@
 \ex@.2326ex
 \def\dddot#1{{\mathop{#1}\limits^{\vbox to-1.4\ex@{\kern-\tw@\ex@
  \hbox{\tenrm...}\vss}}}}
 \makeatother
\thispagestyle{empty}
{\baselineskip0pt
\leftline{\large\baselineskip16pt\sl\vbox to0pt{\hbox{Department of Physics} 
               \hbox{Kyoto University}\vss}}
\rightline{\large\baselineskip16pt\rm\vbox to20pt{\hbox{KUNS-1521} 
\vss}}%
}
\vskip1cm
\begin{center}
{\large{\bf 
Friedmann Limit of a point mass universe (is not Friedmann)
}}\\
{\large --- Gravinational Lens Effect ---} \\
\end{center}
\begin{center}
{\large Norimasa Sugiura\footnote{E-mail:sugiura@tap.scphys.kyoto-u.ac.jp}, 
}\\
{\em Department of Physics,~Kyoto University, 
 Sakyo-ku, Kyoto 606-8502, Japan}
\end{center}
\begin{abstract}
We study the distance-redshift relation in
a universe filled with point particles, and discuss
what the universe looks like when we make the 
number of particles $N$ very large, while fixing the averaged mass
density. 
Using the Raychaudhuri equation and a simple analysis
of the probability of strong lensing effects,
we show that the statistical nature of the amplification is 
independent of $N$, and clarify the appearance of the point particle universe.
\end{abstract}
\pacs{PACS numbers:}

\multicols{2}

\section{Introduction}
Over the last several decades, a great deal of interest has been paid
on the (cumulative) gravitational lensing effect on distant sources
due to inhomogeneites in the
matter distribution of the universe.
This problem has been studied using various methods
\cite{rf:KANTOWSKI,rf:DYERROEDER1,rf:SASAKI,rf:SCHNEIDER,rf:SEF,rf:RAUCH,rf:WCXO,rf:PMM,rf:WCO,rf:TOMITA,rf:HW}.
In some cases, the lens objects can be treated as
point particles.
The point particles may be galaxies for cosmological lenses, or stars for
microlensing events. 
While surveying these papers,
one question arises to us out of purely theoretical interst:
what happens
when we bring the number of particles $N$ very large,
while fixing the mass density?
Does it look like a Friedman--Lema\^{\i}tre(FL) universe, or a
completely different universe?

In this article, we discuss what the universe looks like when we take the 
large $N$ limit of a universe filled with point particles
by studying the distance-redshift relation\footnote{The analysis made in this article is valid for 
any compact object whose mass is contained within its Einstein
radius.}
.

\section{Distance-redshift relation in a point mass universe}
We distribute point particles of the same mass $m$ uniformly
throughout the universe with a mean separation $l$.
We assume that on large scales, the spacetime is described by 
an isotropic homogeneous metric (Robertson--Walker metric).
The energy density parameter $\rho$ is of order of $m/l^3$.
Condier a photon beam which is emitted from a distant source
which we also treat as a point source.
We observe the redshift and the luminosity of this source.
During the propagation, the luminosity of the photon beam 
may be amplified by the gravinational lensing effect.
We can consider two types of lensing effect:
\begin{itemize}
\item Strong lensing effect;
when the beam passes very near to a point particle, it
suffers a strong apmlification.
\item Cumulative weak lensing effect;
the beam does not pass very near to any particle, but travels through
the ``ripples'' of gravitational potential, 
and suffers a weak amplification many times.
\end{itemize}
The cumulative amplification of the weak lensing effect is 
estimated as follws\cite{rf:PG,rf:SACHS}. 

The expansion, $\theta = {1\over 2} {k^a}_{;a}$, of the null geodesic satisfies
\begin{eqnarray}
  {d\theta \over d\lambda} = \theta^2 - |\sigma|^2
\end{eqnarray}
where $\lambda$ is the affine parameter and 
$\sigma$ is the shear;
$|\sigma|^2 = {1\over 2}k_{(a;b)} k^{a;b}-\theta^2$.
We neglected the vorticity.
Since we assume that the beam does not pass 
very near the particles,
the evolution of the shear $\sigma$
is estimated as
\begin{eqnarray}
  {d\sigma \over d\lambda} = - 2 \theta \sigma + {\cal O}(m/l^3) .
\end{eqnarray}
Thus, the change in $\sigma$ during passing by one particle
is approximately given by
\begin{eqnarray}
  \Delta \sigma \sim l(m/l^3).
\end{eqnarray}
Then, the ``random walk'' for distance $L$ results in
\begin{eqnarray}
  \sigma(L) = {\cal O}[ (L/l)^{1/2}l(m/l^3)].
\end{eqnarray}
This leads to
\begin{eqnarray}
  \theta(L) = {\cal O}[L^2 l(m/l^3)^2].
\end{eqnarray}
Since we assume $m/l^3 \sim \rho = 3 \Omega_0 \H0^2 /8\pi$ and
$L\sim \H0^{-1}$ where $\H0$ is the Hubble constant, we obtain
\begin{eqnarray}
\sigma(L)L &\sim& {\cal O}[(l/L)^{1/2}] << 1 ,\label{eq:weak1}\\
\theta(L)L &\sim& {\cal O}[(l/L)] << 1 .\label{eq:weak2}
\end{eqnarray}
In the FL case, on the contrary, the shear term vanishes and the Ricci
focusing term determines the evolution of the expansion;
\begin{eqnarray}
  {d\theta \over d\lambda} = \theta^2 + {\cal O}(m/l^3) \quad .
\end{eqnarray}
This leads to $\theta(L) L \sim 1$.
Thus, the cumulative amplification is negligible in the point mass
universe when the number of paritcles is large enough.
Note that we have assumed that the relation between the affine parameter
and the redshift coincides with that of FL model.
The proof that the difference is negligible
is given in \cite{rf:KANTOWSKI}.

Next, we estimate the probability that the photon beam suffers
the strong lensing effect as follows\footnote{We here do not take
into account of the expansion of the universe, but it would
not change the essential point.
See \cite{rf:PEEBLES} for discussion including the cosmic expansion.}.
We adopt the thin lens approximation.
The Einstein radius of the lens whose mass is $m$ is given
by\cite{rf:PEEBLES} 
\begin{eqnarray}
  \rE = 2\sqrt{mR}
\end{eqnarray}
where $R$ is the distance from the observer to the source
\footnote{
  The discussion in this article holds when the whole mass of the
  object is contained within the Einstein radius.
  However, the Einstein radius for a galaxy is usually smaller than its size.
  The Einstein radius is
  $\sim 20 \sqrt{z}$ Kpc when $m=10^{12}\Msun$,
and
  $\sim 10^5 \sqrt\protect{z\protect}$ times the solar radius
  when $m= \Msun$, 
  where $z$ is the redshift of the lens.
  Thus, a star is usually smaller, and a
  galaxy is larger, than its Einstein radius.
}. 
The magnification factor $A$, the ratio of the flux density in the 
observed image to the flux density in the absence of the lens,
is
\begin{eqnarray}
  A = {1+ 2 \rE^2/b^2 \over (1+4\rE^2/b^2)^{1/2}}
\end{eqnarray}
where $b$ is the impact parameter
\footnote{We do not consider a case there are two (or more) images;
  we assume that we cannot resolve them or the flux of one image
  is much stronger than the ohter. This may be justified from the fact
 that the ratio of flux densities is 7 when $b=\rE$, and becomes
  larger as $b$ increases.
}.
From this expression, we can tell that the source is strongly
amplified (i.e., $A-1$ becomes of order of 10\%) when the impact 
parameter is close to the Einstein radius; $b\sim \rE$.
We cam estimate the probability $P$ that the source
is strongly amplified, 
by considering the probability of hitting any of $N$ discs of radius
$\rE$ when a photon beam travels in a tube of area $S$ (see Fig.1);
\begin{eqnarray}
  P = {\pi\rE^2 N\over S} = {Nm\over SL}\times{\pi\rE^2 L\over m}
 = 4\pi\rho L R
\end{eqnarray}
where $L$ is the distance to the source, and $\rho$ is the mean mass
density of the point mass universe. 
Since $\rho = 3 \Omega_0 \H0^2 /8\pi$ and $L,R \sim z\H0^{-1}$
where $\H0$ is the Hubble constant,
the probability $P$ is much smaller than unity
for sources at low redshifts, or in a low density universe, and
$P\sim 1$ even at rather high redshifts $z\sim1$.

When $P(z)$ is smaller than unity,
part of the beams $1-P(z)$ will reach us without hitting any disc.
The distance (which is estimated from the observed flux) to such sources
is obtained by following the evolution of the flux in an empty
spacetime. That is, we can regard the distance to these sources as
so-called Dyer-Roeder distance\cite{rf:DR2}.
Therefore, when we observe the luminosity and redshift of distant
point sources in the point mass universe,
we would obtain the following:
\begin{itemize}
\item Part of sources $1-P(z)$ follow the distance-redshift
  relation of Dyer-Roeder distance;
\item Other sources $P(z)$ are strongly amplified.
\end{itemize}
The fraction, $1-P(z)$, of the sources which never hit a dics is 
invariant when we change the value of $N$.
Also, the statistical nature of the distribution of amplification 
factor is clearly
independent of $N$ when $P(z)$ is enough smaller than unity.
Therefore, we can say
\begin{itemize}
\item these natures are independent of the number of particles $N$
if the mass density $\rho$ is fixed.
\end{itemize}
Actually, this is the well known fact ``the optical depth of 
gravitational lens is independent of the mass of the lens.''
We here point out that the statement is valid even 
in the large $N$ limit.
That is, even if the universe is filled not with 
stars but with much smaller point particles,
the distance-redshift relation of point sources
satisfies the above features,
as long as we can keep our assumptions, such as geoemtrical 
optics treatment.

The fraction $P(z)$ increases for higher redshifts and
a high density universe.
Then, $P(z)$ becomes larger than unity.
The number of beams which never hit a disc becomes very small,
and multi-scattered events dominate.
However, we expect that the resulting distribution of
luminosity of distant point sources
is insensitive to $N$
as long as the distribution of point particles is random
and the thin lens approximation holds.
This is because we can repeat the same discussion for the 
probability of suffering a next strong lensing event 
after hitting one disc.

\section{Discussion}
The above analysis is valid for any matter 
as long as they are compact enough and 
interact with photons only through gravity.
The lensing objects need not to be galaxies or stars;
they may be sands or elementary (but dark) particles.
Moreover, the thin lens approximation we adopted above
seems to become better as we decrease the mass of lens,
since the ratio of Einstein radius $\rE$ to the means separation $l$
becomes smaller; $l\propto m^{1/3}, \rE\propto m^{1/2}$.
Clearly, the effect of cumulative weak lensing effect becomes 
negligible as we can see from equations (\ref{eq:weak1}) and (\ref{eq:weak2}).
Thus, we conclude that the behaviour of distance-redshift relation 
of the point mas universe 
does not agree with that of a FL model even when
we take the large $N$ limit.

Holz and Wald \cite{rf:HW} studied the lensing effect when the matter
distribution 
of the universe is not homogeneous but the masses are concentrated
into compact objects. 
They commented that the probability distributions of lensing effect are
indistinguishable between the cases where the masses of lenses are
$M=10^{12}\Msun,M=10^{13}\Msun$ and $M=\Msun$.
Their results may support the correctness of the above analysis.
Related with this, one can show that, in their formalism,
the observed ditance-redshift relation
for any point source follows that of a FL
model in the case of uniform (not discrete)
distribution of matter, though they do not give an explicit
statement. We give a rough proof in the appendix.

In Sugiura et al.\cite{rf:SNH}, it was shown that the discreteness of
matter distribution is harmless
when we consider the distance-redshift relation
in a spherically symmetric space.
It suggests that, in the point mass universe,
if we average the luminosity of the sources of the same redshift
over the whole sky and calculate the distace-redshift relation
with the averaged luminosity, it should agree with the FL relation.
Holz and Wald \cite{rf:HW} state that the averaged luminosity of 
the beams agrees with that of a FL model.
These statements justify the result we obtained in \cite{rf:SNH}.

We also notice that, we would obtain the FL relation if
we take the average of sources 
over a region larger than the mean separation
of the interviening lens objects, i.e., over the region which includes
enough strong lensing events.
That is, when the source is much larger than the intervening lens
objects. we can safely calculate its distance using a FL model.
\\
\\
The author would like to thank K.~Ioka and T.~Hamana for useful comments.

\appendix
\section*{}
We show that the distant--redshift relation obtained by the method in
\cite{rf:HW} agrees with that of a FL model in the case of uniform
and continuous density field.

We start from the geodesic deviation equation.
Let $\eta^a$ be the deviation vector, and define matrix $A^{ab}$ by
\begin{eqnarray}
  \eta^a(\la) = A^a_b(\la) {d\eta^b\over d\la}(0)
\end{eqnarray}
where $\la$ is the affine parameter.
Then the geodesic deviation equation is written as
\begin{eqnarray}
  {d^2 {A^a}_b\over d\la^2}= - {R_{cde}}^a k^c k^e {A^d}_b ,
\end{eqnarray}
where $k^a$ is the tangent vector of the null geodesic.
In a Robertson-Walker spacetime, this equation takes the form
\begin{eqnarray}
  {d^2 {A^a}_b\over d\la^2}= - 4\pi \omega^2 \rho {A^a}_b ,
\end{eqnarray}
where $\omega$ is the frequency of the photon
and $\rho$ is the energy density of the universe.
Then, after traveling for small $\Delta \la$,
\begin{eqnarray}\label{dev1}
  {d {A^a}_b\over d\la}(\la + \Delta \la)
  = {d {A^a}_b\over d\la}(\la) - 
     4\pi \omega^2 \rho \Delta \la {A^a}_b(\la) . 
\end{eqnarray}
Consider a ball of radius $R$ whose density is uniform, and 
a bundle of light ray which passes through this ball at a distance $b$ from the
center of the ball. 
By direct calculation, they show that
\begin{eqnarray}\label{dev2}
  {d {A^a}_b\over d\la}(\la + \Delta \la)
  = {d {A^a}_b\over d\la}(\la) 
    - \omega J {A^a}_b(\la) 
\end{eqnarray}
where 
\begin{eqnarray}
  J = 6 M (1-b^2/R^2)^{1/2} /R^2,
\end{eqnarray}
where $M$ is the total mass of the matter inside the ball.
From the relations
\begin{eqnarray}
 \omega \Delta \la = 2 (R^2-b^2)^{1/2} 
\end{eqnarray}
and
\begin{eqnarray}
 \rho &=& \left({4\pi\over 3}R^3\right)^{-1}M ,
\end{eqnarray}
we can see that equation (\ref{dev2}) agrees with (\ref{dev1}).

\begin{figure}[t]
\epsfysize=2.6cm \epsfbox{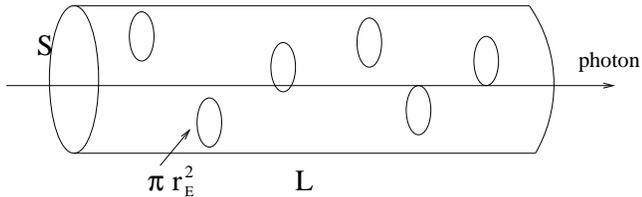}
\caption{Photon beam traveling through a tube filled with point mass
  particles. Each particle is regarded as a disc (or sphere) of radius
  $\rE$.
If the beam hits a disc, it will be strongly amplified.}
\end{figure}
\end{document}